\documentclass[12pt]{article}
\usepackage{amssymb,amsmath,amsfonts}
\usepackage{graphicx}
\usepackage{graphics}
\usepackage{eepic,epsfig}

\begin{document}

\title{Fermionic condensate in de Sitter spacetime}
\author{A. A. Saharian$^1$\thanks{%
E-mail: saharian@ysu.am},\, E. R. Bezerra de Mello $^{2}$\thanks{%
E-mail: emello@fisica.ufpb.br},\, A. S. Kotanjyan$^1$\thanks{%
E-mail: anna.kotanjyan@ysu.am},\, \and T. A. Petrosyan$^1$\thanks{%
E-mail: tigran.petrosyan@ysu.am} \\
\\
\textit{$^1$Department of Physics, Yerevan State University,}\\
\textit{1 Alex Manoogian Street, 0025 Yerevan, Armenia}\vspace{0.3cm} \\
\textit{$^{2}$Departamento de F\'{\i}sica, Universidade Federal da Para\'{\i}%
ba}\\
\textit{58.059-970, Caixa Postal 5.008, Jo\~{a}o Pessoa, PB, Brazil}}
\maketitle

\begin{abstract}
Fermionic condensate is investigated in $(D+1)$-dimensional de Sitter
spacetime by using the cutoff function regularization. In order to fix the
renormalization ambiguity for massive fields an additional condition is
imposed, requiring the condensate to vanish in the infinite mass limit. For
large values of the field mass the condensate decays exponentially in odd
dimensional spacetimes and follows a power law decay in even dimensional
spacetimes. For a massless field the fermionic condensate vanishes for odd
values of the spatial dimension $D$ and is nonzero for even $D$. Depending
on the spatial dimension the fermionic condensate can be either positive or
negative. The change in the sign of the condensate may lead to instabilities
in interacting field theories.
\end{abstract}

\bigskip

Keywords: \textit{fermionic condensate; de Sitter spacetime, Bunch-Davies
vacuum}

\bigskip

\section{Introduction}

De Sitter (dS) spacetime is among the frequently used background geometries
for the investigation of the influence of gravitational field on quantum
matter. In the early stages of studies this interest was motivated by high
symmetry of the corresponding geometry. The dS spacetime is the maximally
symmetric solution of Einstein's equation with a positive cosmological
constant as the only source of gravitational field and because of that a
relatively large number of physical problems can be exactly solved on that
background. This helps to shed light on the effects of gravity on quantum
fields in more complicated geometries. The further increase of the interest
to the investigations of quantum effects on dS bulk was related to the
appearance of the inflationary scenario for the expansion of the early
Universe (for reviews see \cite{Lind90,Mart14}). In most inflationary models
the expansion is described by an approximately dS geometry sourced by the
potential energy of a scalar field (inflaton). A short period of the
corresponding quasi-exponential expansion provides a natural solution to a
number of fine tuning problems of the standard Big Bang model (horizon and
flatness problems, the problem of topological defects, etc.). In addition,
the inflationary scenario leads to an interesting mechanism for the
generation of small inhomogeneities in the energy distribution at the
beginning of the radiation dominated cosmological expansion that seed the
large scale structure of the Universe at late stages. This mechanism is
based on the classicalization of quantum fluctuations of scalar fields by an
inflationary expansion. Its predictions are in good agreement with the
observational data about the temperature anisotropies of the cosmic
microwave background. Those data, in combination with observations of high
redshift supernovae and galaxy clusters indicate that the expansion of the
Universe at the present epoch is well approximated by a model where the
dominant part of the energy content is described by the equation of state
close to the one for a positive cosmological constant. The cosmological
expansion with this type of gravitational source will lead to an
asymptotically dS universe as the future attractor. This shows that the
investigation of physical effects in dS spacetime is also important for the
future of the Universe.

The expectation values of bilinear combinations of quantum fields with
different spins (field squared, energy-momentum tensor) for the Bunch-Davies
vacuum in dS spacetime have been investigated in a large number of papers
(see \cite{Birr82}-\cite{Dolg06} and references therein). In particular, the
Green function and the effective Lagrangian for a spinor field have been
discussed in \cite{Cand75}. The expression for the renormalized vacuum
expectation value (VEV) of the energy-momentum tensor for a spinor field in
4-dimensional dS spacetime is derived in \cite{Mama81} by using the $n$-wave
regularization method. The same result is obtained in \cite{Beze10} by using
the regularization based on a cutoff function. In \cite{Beze10} the
fermionic condensate is investigated as well. The fermionic condensate and
the VEV of the energy-momentum tensor for a spinor field in $(D+1)$%
-dimensional dS spacetime for even values of $D$ have been investigated in
\cite{Beze08} by using the point-splitting regularization technique. The
shifts in the VEVs for spinor fields induced by the toroidal
compactification of a part of spatial dimensions in dS spacetime were
studied in \cite{Beze08,Saha08,Bell13}. Another class of topological effects
caused by the presence of a cosmic string in dS bulk have been discussed in
\cite{Beze10}.

In the present paper we investigate the renormalized fermionic condensate in
$(D+1)$-dimensional dS spacetime for general value of the spatial dimension $%
D$. The regularization procedure will be based on the introduction of a
cutoff function in the corresponding integral representation. In addition to
the VEV of the energy-momentum tensor, the fermionic condensate is an
important local characteristic of the fermionic vacuum. Though the
corresponding operator is local, because of the global nature of the notion
of vacuum, it contains information about global properties of the background
geometry. The fermionic condensate is an important characteristic in quantum
chromodynamics, in the physics of superconductivity and phase transitions,
in models of dynamical mass generation and symmetry breaking. It has been
investigated in various types of physical models, including the ones for
curved backgrounds (see, for example, \cite{Bell11}-\cite{Bell21}).

The paper is organized as follows. In the next section, we describe the
background geometry and present the complete set of fermionic normal modes.
The expression for the fermionic condensate, regularized with the help of
cutoff function, is provided. The extraction of divergences and the
renormalization of the corresponding VEV\ differ for dS spacetimes with even
and odd numbers of spatial dimensions and we describe the respective
procedures in sections \ref{sec:Even} and \ref{sec:Odd}, respectively.
Closed analytic expressions are derived for the renormalized fermionic
condensate in both these cases. In section \ref{sec:Interact} we consider a
model with interacting scalar and fermionic fields where the fermion
condensate determines the effective mass of the scalar field. The main
results are summarized in section \ref{sec:Conc}.

\section{Regularized fermionic condensate in dS spacetime}

\label{sec:RegFC}

We consider a quantum fermionic field $\psi $ on background of $(D+1)$%
-dimensional de Sitter spacetime described by the line element%
\begin{equation}
ds^{2}=dt^{2}-e^{2t/\alpha }\sum_{i=1}^{D}(dz^{i})^{2},  \label{ds2deSit}
\end{equation}%
in planar coordinates $(t,z^{1},\ldots ,z^{D})$. The parameter $\alpha $
determines the Hubble constant and is related to the corresponding positive
cosmological constant $\Lambda $ by the formula $\alpha
^{2}=D(D-1)/(2\Lambda )$. In addition to comoving time coordinate $t$, we
will use the conformal time $\tau $ defined by the relation $\tau =-\alpha
e^{-t/\alpha }$, $-\infty <\tau <0$. In terms of this coordinate, the line
element (\ref{ds2deSit}) takes a conformally flat form with the conformal
factor $(\alpha /\tau )^{2}$. The dynamics of the field in a curved
spacetime is governed by the Dirac equation
\begin{equation}
i\gamma ^{\mu }\left( \partial _{\mu }+\Gamma _{\mu }\right) \psi -m\psi =0\
,  \label{Direq}
\end{equation}%
where $\gamma ^{\mu }=e_{(a)}^{\mu }\gamma ^{(a)}$ are the curved spacetime
Dirac matrices and $\Gamma _{\mu }$ is the spin connection. The vielbein
fields obey the relation $e_{(a)}^{\mu }e_{(b)}^{\nu }\eta ^{ab}=g^{\mu \nu
} $, with $\eta ^{ab}$ being the Minkowski spacetime metric tensor and $%
g_{\mu \nu }=\mathrm{diag}(1,-e^{2t/\alpha },\ldots ,-e^{2t/\alpha })$. The
flat-space Dirac matrices $\gamma ^{(a)}$ are $N\times N$ matrices with $%
N=2^{[(D+1)/2]}$, where the square brackets mean the integer part of the
enclosed expression. In the discussion below these matrices will be taken in
the Dirac representation:
\begin{equation}
\gamma ^{(0)}=\left(
\begin{array}{cc}
1 & 0 \\
0 & -1%
\end{array}%
\right) ,\;\gamma ^{(a)}=\left(
\begin{array}{cc}
0 & \sigma _{a} \\
-\sigma _{a}^{+} & 0%
\end{array}%
\right) ,  \label{gam0l}
\end{equation}%
with $a=1,2,\ldots ,D$ and $(N/2)\times (N/2)$ matrices $\sigma _{a}$. By
using the anticommutation relations for $\gamma ^{(a)}$ one gets $\sigma
_{a}\sigma _{b}^{+}+\sigma _{b}\sigma _{a}^{+}=2\delta _{ab}$. For the
geometry under consideration we can take the vielbein fields in the form $%
e_{\mu }^{(0)}=\delta _{\mu }^{0}$,$\;e_{\mu }^{(a)}=e^{t/\alpha }\delta
_{\mu }^{a}$,$\;a=1,2,\ldots ,D$. The components of the spin connection are
expressed as $\Gamma _{0}=0$,$\;\Gamma _{l}=(e^{t/\alpha }/2\alpha )\gamma
^{(0)}\gamma ^{(l)}$,$\;l=1,2,\ldots ,D$.

The fermionic condensate in the vacuum state $|0\rangle $ is defined as the
VEV $\langle 0|\bar{\psi}\psi |0\rangle =\langle \bar{\psi}\psi \rangle $,
where the Dirac adjoint is expressed as $\bar{\psi}=\psi ^{\dagger }\gamma
^{(0)}$. In the discussion below we will assume that the state $|0\rangle $
corresponds to the maximally symmetric Bunch-Davies vacuum. Note that the
maximal symmetry does not uniquely define the vacuum state. As it has been
discussed in \cite{Alle85}, in dS spacetime there is a one-complex-parameter
family of maximally symmetric states. Among those states the Bunch-Davies
vacuum is singled out as the only state having the Hadamard structure of
singularities.

Given the complete set of solutions to the equation (\ref{Direq}), denoted
here as $\{\psi _{\beta }^{(+)},\psi _{\beta }^{(-)}\}$, the fermion
condensate is written as the mode-sum
\begin{equation}
\langle \bar{\psi}\psi \rangle =\frac{1}{2}\sum_{\beta }\left( \bar{\psi}%
_{\beta }^{(-)}\psi _{\beta }^{(-)}-\bar{\psi}_{\beta }^{(+)}\psi _{\beta
}^{(+)}\right) .  \label{Cond}
\end{equation}%
Here, $\psi _{\beta }^{(+)}$ and $\psi _{\beta }^{(-)}$ are the analogs of
the positive and negative energy mode functions in the Minkowski bulk and
the collective index $\beta $ presents the set of quantum numbers. In (\ref%
{Cond}), the symbol $\sum_{\beta }$ is understood as a summation over the
discrete quantum numbers and an integration over the continuous ones. In the
problem under consideration the mode functions are specified by the momentum
$\mathbf{k}=(k^{1},\ldots ,k^{D})$ and by the quantum number $\sigma $
taking the values $\sigma =1,\ldots ,N/2$ (hence, $\beta =(\mathbf{k},\sigma
)$). They are given by the expressions%
\begin{eqnarray}
\psi _{\beta }^{(+)} &=&C(k)\eta ^{(D+1)/2}e^{i\mathbf{k}\cdot \mathbf{r}%
}\left(
\begin{array}{c}
H_{1/2-i\alpha m}^{(1)}(k\eta )w_{\sigma }^{(+)} \\
-i(\mathbf{n}\cdot \boldsymbol{\sigma })H_{-1/2-i\alpha m}^{(1)}(k\eta
)w_{\sigma }^{(+)}%
\end{array}%
\right) ,  \notag \\
\psi _{\beta }^{(-)} &=&C(k)\eta ^{(D+1)/2}e^{i\mathbf{k}\cdot \mathbf{r}%
}\left(
\begin{array}{c}
-i(\mathbf{n}\cdot \boldsymbol{\sigma })H_{-1/2+i\alpha m}^{(2)}(k\eta
)w_{\sigma }^{(-)} \\
H_{1/2+i\alpha m}^{(2)}(k\eta )w_{\sigma }^{(-)}%
\end{array}%
\right) ,  \label{psibet}
\end{eqnarray}%
where $\eta =-\tau $, $k=|\mathbf{k}|$, $\mathbf{n}=\mathbf{k}/k$, $\mathbf{k%
}\cdot \mathbf{r}=\sum_{i=1}^{D}k^{i}z^{i}$, $H_{\nu }^{(1,2)}(z)$ are the
Hankel functions, and $\boldsymbol{\sigma }=(\sigma _{1},\sigma _{2},\ldots
,\sigma _{D})$. In (\ref{psibet}), the one-column matrices $w_{\sigma
}^{(\pm )}$ have $N/2$ rows and the elements $w_{\sigma l}^{(+)}=\delta
_{\sigma l}$, $w_{\sigma l}^{(-)}=i\delta _{\sigma l}$. The normalization
coefficient $C(k)$ is expressed as
\begin{equation}
C(k)=\frac{\sqrt{k}e^{\pi \alpha m/2}}{2^{D/2+1}\pi ^{(D-1)/2}\alpha ^{D/2}}.
\label{Cbet}
\end{equation}%
Similar mode functions in locally dS spacetime with a toroidally
compactified subspace are presented in \cite{Beze08}. The mode functions for
Dirac fermions in 4-dimensional dS spacetime have also been considered in
\cite{Cota02}. For a massless field, by taking into account that $%
H_{1/2}^{(1)}(x)=-i\sqrt{2/\pi x}e^{ix}$, we get the conformal relation $%
\psi _{\beta }^{(\pm )}=(\eta /\alpha )^{D/2}\psi _{\mathrm{(M)}\beta
}^{(\pm )}$ with the corresponding modes in Minkowski spacetime.

Substituting the normal modes (\ref{psibet}) in (\ref{Cond}), for the
fermionic condensate we find
\begin{eqnarray}
\langle \bar{\psi}\psi \rangle &=&\frac{\eta ^{D+1}e^{\pi \alpha m}N}{%
2^{D+2}\pi ^{D/2-1}\Gamma (D/2)\alpha ^{D}}\int_{0}^{\infty }dk\,k^{D}
\notag \\
&&\times \left[ H_{-1/2-i\alpha m}^{(1)}(k\eta )H_{-1/2+i\alpha
m}^{(2)}(k\eta )-H_{1/2-i\alpha m}^{(1)}(k\eta )H_{1/2+i\alpha
m}^{(2)}(k\eta )\right] .  \label{FC0}
\end{eqnarray}%
The expression on the right-hand side is divergent and some renormalization
procedure is necessary. Introducing the Macdonald function instead of the
Hankel function, the formula (\ref{FC0}) is rewritten as
\begin{eqnarray}
\langle \bar{\psi}\psi \rangle &=&\frac{2^{-D}\alpha ^{-D}\eta ^{D+1}N}{i\pi
^{D/2+1}\Gamma (D/2)}\left( \partial _{\eta }+\frac{1-2im\alpha }{\eta }%
\right)  \notag \\
&&\times \int_{0}^{\infty }dk\,k^{D-1}K_{1/2-im\alpha }(ik\eta
)K_{1/2-im\alpha }(-ik\eta ).  \label{FC1}
\end{eqnarray}%
In deriving this representation we have used the relation%
\begin{equation}
K_{\nu }(y)K_{\nu -1}(-y)-K_{\nu }(-y)K_{\nu -1}(y)=\left( \partial _{y}+%
\frac{2\nu }{y}\right) K_{\nu }(y)K_{\nu }(-y),  \label{RelMac}
\end{equation}%
with $y=ix$ and $\nu =1/2-im\alpha $. This relation directly follows from
the recurrence relations for the Macdonald function.

In order to obtain an alternative integral representation of the fermionic
condensate, for the product of the Macdonald functions (\ref{FC1}) we use
the formula \cite{Wats66}
\begin{equation}
K_{\nu }(ik\eta )K_{\nu }(-ik\eta )=\int_{0}^{\infty }dy\,\cosh (2\nu
y)\int_{0}^{\infty }\frac{du}{u}\,\exp \left[ -2(k\eta \sinh y)^{2}u-\frac{1%
}{2u}\right] .  \label{RelMac2}
\end{equation}%
Substituting this into (\ref{FC1}), we first integrate over $k$. Then,
instead of $u$ we introduce a new integration variable $x=1/(u\eta ^{2}\sinh
^{2}y)$. After changing the order of the integrations, the integral over $y$
is expressed in terms of the Macdonald function and we find
\begin{equation}
\langle \bar{\psi}\psi \rangle =-\frac{i\alpha ^{-D}\eta ^{D+1}N}{2(2\pi
)^{D/2+1}}\left( \partial _{\eta }+\frac{1-2im\alpha }{\eta }\right)
\int_{0}^{\infty }dx\,\,x^{D/2-1}e^{x\eta ^{2}}K_{1/2-im\alpha }(x\eta ^{2}).
\label{FC2}
\end{equation}%
Using the relation
\begin{equation}
\left( \eta \partial _{\eta }+2\nu \right) e^{x\eta ^{2}}K_{\nu }(x\eta
^{2})=2x\eta ^{2}e^{x\eta ^{2}}\left[ K_{\nu }(x\eta ^{2})-K_{\nu -1}(x\eta
^{2})\right] ,  \label{RelMac3}
\end{equation}%
the condensate can also be presented in the form%
\begin{equation}
\langle \bar{\psi}\psi \rangle =\frac{2\alpha ^{-D}N}{(2\pi )^{D/2+1}}%
\int_{0}^{\infty }dx\,\,x^{D/2}e^{x}{\mathrm{Im}}\left[ K_{1/2-im\alpha }(x)%
\right] .  \label{FC3}
\end{equation}%
The integral in the right-hand side diverges in the upper limit.

For the further evaluation an explicit regularization scheme should be used.
As such a scheme we will introduce an exponential cutoff function $e^{-sx}$,
$s>0$, in the integrand of (\ref{FC3}) with the regularized expression%
\begin{equation}
\langle \bar{\psi}\psi \rangle ^{(s)}=\frac{2\alpha ^{-D}N}{(2\pi )^{D/2+1}}%
\int_{0}^{\infty }dx\,\,x^{D/2}e^{(1-s)x}{\mathrm{Im}}\left[ K_{1/2-im\alpha
}(x)\right] .  \label{FC4}
\end{equation}%
The limit $s\rightarrow 0$ should be taken at the and of calculations.

In (\ref{FC4}), the integral over $x$ is explicitly evaluated in terms of
the associated Legendre function (see \cite{Prud86}) and we find%
\begin{equation}
\langle \bar{\psi}\psi \rangle ^{(s)}=\frac{\alpha ^{-D}N}{(2\pi )^{(D+1)/2}}%
{\mathrm{Im}}\left[ \Gamma \left( \mu +im\alpha \right) \Gamma \left( \mu
+1-im\alpha \right) \frac{P_{-im\alpha }^{-\mu }(-\gamma )}{(1-\gamma
^{2})^{\mu /2}}\right] ,  \label{FC5s}
\end{equation}%
with $\gamma =1-s$ and
\begin{equation}
\mu =\frac{D+1}{2}.  \label{mu}
\end{equation}%
For the product of the gamma functions in this formula one has%
\begin{equation}
\Gamma \left( \mu +im\alpha \right) \Gamma \left( \mu +1-im\alpha \right)
=B_{D}(m\alpha )\left( \mu -im\alpha \right) \prod_{l=1}^{[D/2]}\left[
\left( \mu -l\right) ^{2}+m^{2}\alpha ^{2}\right] ,  \label{GamProd}
\end{equation}%
where $[D/2]$ stands for the integer part of $D/2$, and the function%
\begin{equation}
B_{D}(x)=\left\{
\begin{array}{ll}
\pi x/\sinh (\pi x), & \text{for odd }D, \\
\pi /\cosh (\pi x), & \text{for even }D,%
\end{array}%
\right.  \label{BD}
\end{equation}%
is introduced. Now we want to expand the regularized fermionic condensate in
powers of $s$. The further discussion should be developed for even and odd
values $D$ separately.

\section{Condensate in even dimensional spacetimes}

\label{sec:Even}

First we consider odd values of the spatial dimension $D$. In this case $\mu
$ is an integer and the corresponding Legendre function in (\ref{FC5s}) is
expressed in terms of the hypergeometric function as follows:%
\begin{equation}
P_{-im\alpha }^{-\mu }(-\gamma )=\frac{\Gamma (1-im\alpha -\mu )}{\Gamma
(1-im\alpha +\mu )}(1-\gamma ^{2})^{\mu /2}\partial _{\gamma }^{\mu }F\left(
im\alpha ,1-im\alpha ;1;\frac{1+\gamma }{2}\right) .  \label{LegHyp}
\end{equation}%
Substituting this into the expression for the regularized fermionic
condensate, we get
\begin{equation}
\langle \bar{\psi}\psi \rangle ^{(s)}=-\frac{\pi N\alpha ^{-D}}{(2\pi )^{\mu
}\sinh (\pi m\alpha )}{\mathrm{Re}}\left[ \partial _{s}^{\mu }F(im\alpha
,1-im\alpha ;1;1-s/2)\right] .  \label{FCodd1}
\end{equation}%
The expansion of the right-hand side of this expression is given by the
formula \cite{Abra}%
\begin{equation}
F(im\alpha ,1-im\alpha ;1;1-s/2)=\frac{i}{\pi }\sinh (\pi m\alpha
)\sum_{n=0}^{\infty }a_{n}\left[ b_{n}-\ln (s/2)\right] (s/2)^{n},
\label{Fexp}
\end{equation}%
for the hypergeometric function. In this formula,%
\begin{eqnarray}
a_{n} &=&\frac{(im\alpha )_{n}(1-im\alpha )_{n}}{(n!)^{2}},  \notag \\
b_{n} &=&2\Psi (n+1)-\Psi (n+im\alpha )-\Psi (n+1-im\alpha ),  \label{anbn}
\end{eqnarray}%
where $(c)_{n}$ is Pochhammer's symbol, $\Psi (x)=\Gamma ^{\prime
}(x)/\Gamma (x)$ is the digamma function (here we use the notation $\Psi (x)$
for the digamma function instead of the standard one $\psi (x)$ in order to
avoid the confusion with the fermion field $\psi $). With the use of (\ref%
{Fexp}), we have the following expansion%
\begin{eqnarray}
\langle \bar{\psi}\psi \rangle ^{(s)} &=&\frac{\alpha ^{-D}N}{(4\pi )^{\mu }}%
\left[ \sum_{l=1}^{\mu }C_{l}^{\mu }(-1)^{l}(l-1)!\sum_{n=1}^{l}\frac{{%
\mathrm{Im}}(a_{\mu -n})}{(s/2)^{n}}(l-n+1)_{\mu -l}-\mu !{\mathrm{Im}}%
(a_{\mu })\ln (s/2)\right.   \notag \\
&&\left. +\sum_{l=1}^{\mu }C_{l}^{\mu }(-1)^{l}(l-1)!{\mathrm{Im}}(a_{\mu
})(l+1)_{\mu -l}+\mu !{\mathrm{Im}}(a_{\mu }b_{\mu })+\cdots \right] ,
\label{FCodd2}
\end{eqnarray}%
where $C_{l}^{\mu }$ are the binomial coefficients and the dots stand for
the terms which vanish in the limit $s\rightarrow 0$. As it is seen from (%
\ref{FCodd2}), we have the power-law divergent terms, logarithmically
divergent term, and the finite part. Note that the coefficients (\ref{anbn})
can also be written in the form%
\begin{eqnarray}
a_{n} &=&\frac{im\alpha (n!)^{-2}}{n+im\alpha }\prod_{l=1}^{n}(l^{2}+m^{2}%
\alpha ^{2}),\;a_{0}=1,  \notag \\
b_{n} &=&2\Psi (n+1)-2{\mathrm{Re}}[\Psi (n+im\alpha )]-\frac{1}{n-im\alpha }%
.  \label{anbn2}
\end{eqnarray}

\begin{figure}[tbph]
\begin{center}
\epsfig{figure=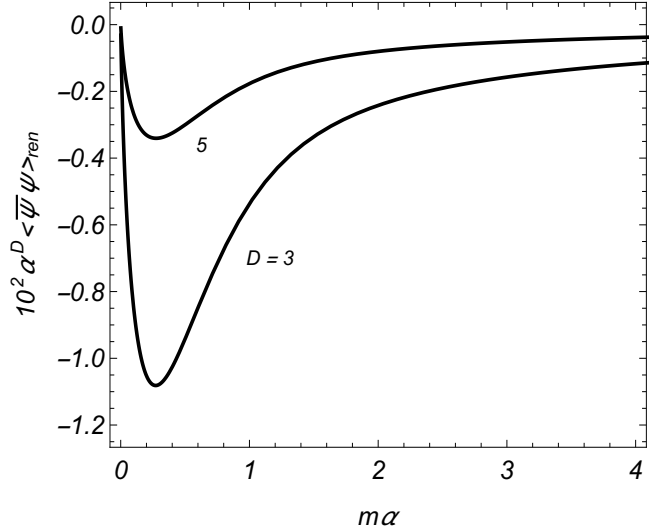,width=8.5cm,height=7.cm}
\end{center}
\caption{Fermionic condensate versus $m\protect\alpha $ for the spatial
dimensions $D=3$ and $D=5$.}
\label{fig1}
\end{figure}

On the basis of the expansion (\ref{FCodd2}), taking into account the finite
renormalization terms, the renormalized fermionic condensate is written in
the form
\begin{eqnarray}
\langle \bar{\psi}\psi \rangle _{\mathrm{ren}} &=&-\frac{\alpha ^{-D}N}{%
(4\pi )^{\mu }}\frac{(m\alpha )^{2\mu -1}}{\Gamma (\mu )}\Big\{%
\sum_{l=1}^{\mu -1}\frac{f_{l}}{(m\alpha )^{2l}}  \notag \\
&&+2\left\{ {\mathrm{Re}}[\Psi \left( im\alpha \right) ]-\ln (m\alpha
)\right\} \prod_{l=1}^{\mu -1}\left( 1+\frac{l^{2}}{m^{2}\alpha ^{2}}\right) %
\Big\},  \label{FCoddRen}
\end{eqnarray}%
where we have used the relation ${\mathrm{Re}}[\Psi \left( 1+im\alpha
\right) ]={\mathrm{Re}}[\Psi \left( im\alpha \right) ]$ which directly
follows from the formula $\Psi \left( 1+z\right) =\Psi \left( z\right) +1/z$
for the digamma function (see \cite{Abra}). In (\ref{FCoddRen}), the
coefficients $f_{l}$ should be fixed by an additional renormalization
condition (for a discussion of ambiguities in the renormalization of the
expectation value of the energy-momentum tensor in the Hadamard
renormalization procedure for general number of spatial dimensions see \cite%
{Deca08}). As such a condition we require that $\langle \bar{\psi}\psi
\rangle _{\mathrm{ren}}\rightarrow 0$ in the limit $m\rightarrow \infty $.
By using the expansion%
\begin{equation}
{\mathrm{Re}}[\Psi \left( im\alpha \right) ]=\ln (m\alpha
)+\sum_{n=1}^{\infty }\frac{(-1)^{n-1}B_{2n}}{2n(m\alpha )^{2n}},
\label{psiExp}
\end{equation}%
with $B_{2n}$ being the Bernoulli coefficients, and requiring the
cancellation of the terms in (\ref{FCoddRen}) with positive powers of the
mass, we find%
\begin{equation}
\sum_{n=1}^{\mu -1}\frac{(-1)^{n}B_{2n}}{nx^{n}}\prod_{l=1}^{\mu -1}\left( 1+%
\frac{l^{2}}{x}\right) =-\sum_{l=1}^{\mu -1}\frac{f_{l}}{x^{l}}+\cdots .
\label{bldef}
\end{equation}%
This relation defines the values of the coefficients $f_{l}$ in the
expression (\ref{FCoddRen}) for the renormalized fermionic condensate. In
particular, one has $f_{1}=-1/6$ for $D=3$, $f_{1}=-1/6$, $f_{2}=-17/20$ for
$D=5$, and $f_{1}=-1/6$, $f_{2}=-47/20$, $f_{3}=-5297/630$ for $D=7$. In the
cases $D=3$ and $D=5$ from (\ref{FCoddRen}) one finds%
\begin{eqnarray}
\langle \bar{\psi}\psi \rangle _{\mathrm{ren}} &=&\frac{m}{2\pi ^{2}\alpha
^{2}}\left\{ \left( \ln (m\alpha )-{\mathrm{Re}}[\Psi \left( im\alpha
\right) ]\right) (m^{2}\alpha ^{2}+1)+\frac{1}{12}\right\} ,  \notag \\
\langle \bar{\psi}\psi \rangle _{\mathrm{ren}} &=&\frac{m}{8\pi ^{3}\alpha
^{4}}\Big\{\frac{m^{2}\alpha ^{2}}{12}+\frac{17}{40}+\left( \ln (m\alpha )-{%
\mathrm{Re}}[\Psi \left( im\alpha \right) ]\right) (m^{2}\alpha
^{2}+1)(m^{2}\alpha ^{2}+4)\Big\}.  \label{FCoddD35}
\end{eqnarray}%
For $D=3$ the result (\ref{FCoddD35}) coincides with the corresponding
expression obtained previously in \cite{Beze10}. For a massless field the
renormalized fermionic condensate vanishes. For large masses, $m\alpha \gg 1$%
, the condensate behaves as $\alpha ^{-D}/(m\alpha )$. In figure \ref{fig1}
we have plotted the fermionic condensate as a function of $m\alpha $ for $%
D=3 $ and $D=5$. In these cases the fermionic condensate is negative for
massive fields.

\section{Fermionic condensate in odd dimensional spacetime}

\label{sec:Odd}

In the renormalization procedure we need the expansion of the expression on
the right-hand side of (\ref{FC5s}) near the point $\gamma =1$. For even
values of $D$, this expansion for the associated Legendre function directly
follows from the formula
\begin{eqnarray}
\frac{P_{-im\alpha }^{-\mu }(-\gamma )}{(1-\gamma ^{2})^{\mu /2}} &=&\frac{%
\left( 2-s\right) ^{-\mu }\sinh (\pi m\alpha )}{i\Gamma (1+\mu )\sin (\mu
\pi )}F(im\alpha ,1-im\alpha ;1+\mu ;s/2)  \notag \\
&&+\frac{1}{\pi }\sin [\pi (im\alpha +\mu )]\frac{\Gamma (1-im\alpha -\mu
)\Gamma (\mu )}{\Gamma (1-im\alpha +\mu )s^{\mu }}F(im\alpha ,1-im\alpha
;1-\mu ;s/2).  \label{LegEven}
\end{eqnarray}%
The standard definition of the hypergeometric function in terms of the
series over $s$ provides the required expansion. For the case under
consideration $\mu $ is a half-integer and, hence, the second term on the
right-hand side of (\ref{LegEven}) does not contribute to the finite part,
whereas the first term is finite in the limit $s\rightarrow 0$. Substituting
expression (\ref{LegEven}) into formula (\ref{FC5s}) and using the expansion
for the hypergeometric function in the second term on the right-hand side,
we find the following expansion for the regularized fermionic condensate%
\begin{eqnarray}
\langle \bar{\psi}\psi \rangle ^{(s)} &=&\frac{N\Gamma ((1-D)/2)}{(4\pi
)^{\mu }\alpha ^{D}}\bigg\{\frac{\Gamma (\mu )}{m\alpha }\sum_{n=1}^{[\mu ]}%
\frac{n(s/2)^{n-\mu }}{\Gamma (1-\mu +n)n!}\prod_{l=0}^{n-1}\left(
l^{2}+m^{2}\alpha ^{2}\right)  \notag \\
&&-\tanh (\pi m\alpha )\prod_{l=0}^{D/2-1}\left[ \left( l+1/2\right)
^{2}+m^{2}\alpha ^{2}\right] +\cdots \bigg\},  \label{FCEven}
\end{eqnarray}%
where, as before, the dots stand for the terms which vanish in the limit $%
s\rightarrow 0$.

\begin{figure}[tbph]
\begin{center}
\epsfig{figure=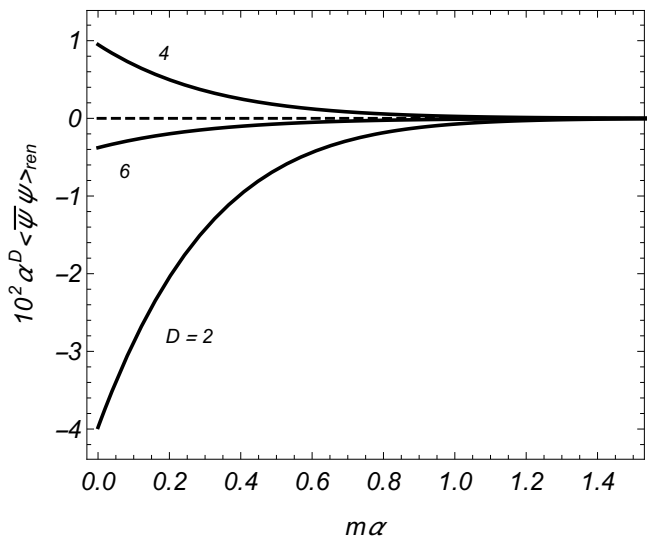,width=8.5cm,height=7.cm}
\end{center}
\caption{Fermionic condensate as a function of $m\protect\alpha $ for $%
D=2,4,6$.}
\label{fig2}
\end{figure}

From formula (\ref{FCEven}) we find the following expression for the
renormalized fermionic condensate:%
\begin{equation}
\langle \bar{\psi}\psi \rangle _{\mathrm{ren}}=-\frac{Nm^{D}}{(4\pi )^{\mu }}%
\Gamma \left( \frac{1-D}{2}\right) \left\{ \sum_{l=0}^{D/2}\frac{c_{l}}{%
(m\alpha )^{2l}}+\tanh (\pi m\alpha )\prod_{l=1}^{D/2}\left[ 1+\left( \frac{%
l-1/2}{m\alpha }\right) ^{2}\right] \right\} ,  \label{FCevenRen}
\end{equation}%
where the coefficients $c_{l}$ are determined from the renormalization
condition $\langle \bar{\psi}\psi \rangle _{\mathrm{ren}}\rightarrow 0$ for $%
m\rightarrow \infty $. From this condition it follows that%
\begin{equation}
\sum_{l=0}^{D/2}\frac{c_{l}}{(m\alpha )^{2l}}=-\prod_{l=1}^{D/2}\left[
1+\left( \frac{l-1/2}{m\alpha }\right) ^{2}\right] .  \label{cl}
\end{equation}%
This leads to the following formula for the fermionic condensate%
\begin{equation}
\langle \bar{\psi}\psi \rangle _{\mathrm{ren}}=\frac{(-1)^{D/2}(4\pi
)^{(1-D)/2}N\alpha ^{-D}}{2\Gamma ((D+1)/2)\left( e^{2\pi m\alpha }+1\right)
}\prod_{l=1}^{D/2}\left[ m^{2}\alpha ^{2}+\left( l-1/2\right) ^{2}\right] .
\label{FCevenRen1}
\end{equation}%
This expression coincides with the result obtained in \cite{Beze08} by using
the point-splitting procedure and the adiabatic subtraction. Hence, we have
shown that the different renormalization schemes give the same result for
the renormalized fermionic condensate. The sign of the fermionic condensate (%
\ref{FCevenRen1}) coincides with the sign of $(-1)^{D/2}$. For large values
of the mass, $m\alpha \gg 1$, the fermionic condensate (\ref{FCevenRen1}) is
suppressed by the factor $m^{D}\alpha ^{D}e^{-2\pi m\alpha }$. Unlike to the
case of odd $D$, in even number of spatial dimensions the fermionic
condensate for a massless field differs from zero (see also \cite{Beze08}):%
\begin{equation}
\langle \bar{\psi}\psi \rangle _{\mathrm{ren}}=\frac{(-1)^{D/2}N\Gamma
((D+1)/2)}{(4\pi )^{(D+1)/2}\alpha ^{D}}.  \label{FCm0}
\end{equation}%
In figure \ref{fig2} the dependence of the fermionic condensate on $m\alpha $
is presented for several values of the spatial dimension.

\section{Interacting scalar and fermion fields}

\label{sec:Interact}

Nonzero fermionic condensate can be of considerable importance in
interacting field theories. As an example, here we consider a system of
interacting fermionic and scalar fields described by the Lagrangian density
\begin{equation}
\mathcal{L}=\frac{1}{2}g^{\mu \nu }\partial _{\mu }\varphi \partial _{\nu
}\varphi -\frac{1}{2}M^{2}\varphi ^{2}+\frac{i}{2}[\bar{\psi}\gamma ^{\mu
}\nabla _{\mu }\psi -(\nabla _{\mu }\bar{\psi})\gamma ^{\mu }\psi ]-m\bar{%
\psi}\psi -\lambda \varphi ^{2}\bar{\psi}\psi ,  \label{Lag}
\end{equation}%
with the coupling constant $\lambda $ having the dimension (length)$^{D-2}$.
The corresponding field equations read%
\begin{eqnarray}
(\square +M^{2}+2\lambda \bar{\psi}\psi )\varphi &=&0,  \notag \\
(i\gamma ^{\mu }\nabla _{\mu }-m-\lambda \varphi ^{2})\psi &=&0,\
\label{FeqSys}
\end{eqnarray}%
where $\square $ stands for d'Alembert operator for scalar fields.

Assume that the field $\psi $ is quantized and the field $\varphi $ is a
classical field. If $\langle \bar{\psi}\psi \rangle _{\mathrm{ren}}$ is the
renormalized fermion condensate, then the classical dynamics of the scalar
field is described by the equation%
\begin{equation}
(\square +M^{2}+2\lambda \langle \bar{\psi}\psi \rangle _{\mathrm{ren}%
})\varphi =0.  \label{phiClasEq}
\end{equation}%
As it is seen, the effect of of the interaction of the scalar field with the
fluctuations of the fermionic field is equivalent to the change of the mass
term. For a general background the effective mass depends on the spacetime
point. In the case of dS bulk the fermion condensate is constant and the
interaction leads to a constant shift in the squared mass term for the
scalar field. In general, this shift can be negative and under the condition
$M^{2}+2\lambda \langle \bar{\psi}\psi \rangle _{\mathrm{ren}}<0$ the
effective mass becomes tachyonic. The tachyonic mass may lead to an
instability of the corresponding field theory (for instabilities in
interacting scalar field theories induced by background geometry, nontrivial
topology and boundaries see \cite{Ford80}). Note that, in a similar way, the
quantum fluctuations of the scalar field lead to the correction of the
fermionic mass term in the form $\lambda \langle \varphi ^{2}\rangle _{%
\mathrm{ren}}$. In a more general case of a scalar field with the potential $%
V(\varphi )$, the interaction with the vacuum fluctuations of a fermionic
field leads to the correction with the effective potential $V_{\mathrm{eff}%
}(\varphi )=V(\varphi )+\lambda \langle \bar{\psi}\psi \rangle _{\mathrm{ren}%
}\varphi ^{2}$. In particular, this type of correction to the inflaton
potential can have important consequences in the inflationary scenario.

Similar to the case of the system of interacting fermion and scalar fields,
the nonzero fermionic condensate leads to the shift of the fermion effective
mass in the Nambu-Jona-Lasinio type models. These models contain four
fermion interaction term $g(\bar{\psi}\psi )^{2}$ in the Lagrangian density,
with $g$ being the four fermion coupling constant. They were applied to
describe the dynamical symmetry breaking in electroweak theory and quantum
chromodynamics (for symmetry breaking in the Nambu-Jona-Lasinio model in
curved spacetime see, for example, \cite{Eliz94b,Inag97}). The corresponding
effective mass for a fermion field becomes $m-2g\langle \bar{\psi}\psi
\rangle _{\mathrm{ren}}$. Again, we see that, depending on the fermionic
condensate, the effective mass may become negative.

\section{Conclusion}

\label{sec:Conc}

In the present paper we have investigated the fermionic condensate for a
massive spinor field in dS spacetime in an arbitrary number of spatial
dimensions. In Section \ref{sec:RegFC}, an expression for the corresponding
regularized quantity is derived assuming that the field is prepared in the
Bunch-Davies vacuum state. The renormalization procedure for even and odd
dimensional spacetimes is considered separately. In even dimensional dS
spacetime the renormalized fermionic condensate is given by expression (\ref%
{FCoddRen}), where the coefficients are obtained from the condition of
vanishing the condensate in the limit $m\rightarrow \infty $. These
coefficients are defined by the relation (\ref{bldef}). For large values of
the field mass, the condensate decays as $1/(m\alpha )$ and it vanishes for
a massless field field. In odd dimensional dS spacetime, for the
renormalized fermionic condensate we derived the formula (\ref{FCevenRen}),
with the coefficients $c_{l}$ defined from the relation (\ref{cl}). In this
case, for large values of the mass the fermionic condensate decays
exponentially. Unlike the case of even dimensions, for a massless field the
condensate does not vanish.

Another vacuum state in dS spacetime is the hyperbolic vacuum \cite{Pfau82}-%
\cite{Saha21b}. It is naturally realized by the normal modes of quantum
fields in the coordinate system foliating the spacetime by spatial sections
with constant negative curvature. Unlike to the Bunch-Davies vacuum, the
hyperbolic vacuum is not maximally symmetric and the corresponding fermionic
condensate will depend on time. This feature has been demonstrated for the
expectation values of the field squared and energy-momentum tensor in the
case of a scalar field (see \cite{Pfau82,Saha21b}). For a massless fermionic
field we expect that the difference in the fermionic condensates for the
Bunch-Davies and hyperbolic vacua will decay at late stages of the expansion
like $1/t^{D}$. This is in agreement with the general result in accordance
of which the Bunch-Davies vacuum is a future attractor for relatively large
class of states in dS spacetime. Note that the renormalization of the
fermionic condensate for the hyperbolic vacuum is reduced to the
renormalization for the Bunch-Davies vacuum and the difference in the
corresponding VEVs is finite.

In interacting field-theoretical models (self-interacting fermionic field,
fermionic fields interacting with scalar or vector fields) the formation of
nonzero fermionic condensate may lead to phase transitions. We have
considered two examples. The first one presents a system of scalar and
fermionic fields with the interaction Lagrangian density proportional to $%
\varphi ^{2}\bar{\psi}\psi $ and the second one corresponds to the
Nambu-Jona-Lasinio type model with the self interaction $(\bar{\psi}\psi
)^{2}$. Depending on the value and sign of the condensate, the effective
mass squared may become negative. Scalar-fermionic models with the
interaction $\mathrm{const}\cdot \bar{\psi}\psi g^{\mu \nu }\partial _{\mu
}\varphi \partial _{\nu }\varphi $ have also been considered in the
literature. In this type of models the nonzero condensate may lead to the
change of the sign of the kinetic term for the scalar field (ghost field).

\section*{Acknowledgments}

A.A.S. and A.S.K. were supported by the grant No. 20RF-059 of the Committee
of Science of the Ministry of Education, Science, Culture and Sport RA.
E.R.B.M. is partially supported by CNPq under Grant no. 301.783/2019-3.
T.A.P. was supported by the Committee of Science of the Ministry of
Education, Science, Culture and Sport RA in the frames of the research
project No. 20AA-1C005.

\end{document}